\documentclass[12pt,preprint]{aastex}

\begin{document}

\title{The Stellar Halo in the Large Magellanic Cloud:
Mass, Luminosity, and Microlensing Predictions}

\author{David R.~Alves }

\affil{Columbia Astrophysics Laboratory, 
550 W. 120th St., New York, NY, USA}

\email{alves@astro.columbia.edu}

\begin{abstract}
Recently obtained kinematic data has shown that
the Large Magellanic Cloud~(LMC) 
possesses an old stellar halo.
In order to further characterize
the properties of this halo, 
parametric King models are fit 
to the surface density of RR Lyrae stars.
Using data from both 
the MACHO and OGLE~II microlensing surveys, 
the model fits 
yield the center of their distribution at
$\alpha$ = $5^{\rm h} \ 21.1 \pm 0.8\,^{\rm m}$,
$\delta$ = $-69^{\rm \circ} \ 45 \pm 6\arcmin$ (J2000)
and a core radius of $1.42 \pm 0.12$ kpc.
As a check
the halo model is compared 
with RR Lyrae star counts in fields near the LMC's periphery
previously surveyed with photographic plates.
These data, however, require a cautious
interpretation.
Several topics regarding the LMC stellar halo are discussed.
First,
the properties of the halo imply
a global mass-to-light ratio of $M/L_V = 5.3 \pm 2.1$
and a total mass of $1.6 \pm 0.6 \times 10^{10} M_{\odot}$
for the LMC
in good agreement with estimates based on the rotation curve.
Second,
although the LMC's disk and halo are kinematically distinct,
the shape of the surface density profile of the halo
is remarkably similar to that of the young disk.  
For example, the best-fit exponential 
scale length for the RR Lyrae stars is $1.47 \pm 0.08$~kpc,
which compares to 1.46~kpc for the LMC's blue light.
In the Galaxy,
the halo and disk do not resemble each other like this.
Finally,
a local maximum in the LMC's microlensing optical depth 
due to halo-on-disk stellar self-lensing is predicted.
For the parameters of the stellar halo obtained,
this maximum is located near MACHO events LMC-4 and LMC-23,
and is large enough to possibly account for these two events,
but not for all of the observed microlensing.
\end{abstract}


\section{ Introduction }

The first strong evidence for the existence
of a kinematically hot and metal-poor
stellar population in the Large Magellanic Cloud (LMC)
has recently been obtained at
the European Southern Obervatory's {\it Very Large Telescope}
(Minniti et al.~2003).  
The radial velocities derived from the spectra of 43 RR Lyrae stars 
imply a velocity dispersion of 53~$\pm$~10~km~s$^{-1}$,
which is larger than that of any other population in the LMC
(e.g., Hughes et al.~1991).
For comparison, the radial velocities of carbon stars, which
represent the bulk population of the LMC disk,
have a dispersion of $20.5 \pm 0.5$~km~s$^{-1}$
(van der Marel et al.~2002).  Simple equilibrium models
of the carbon-star radial velocities
suggest that the LMC disk is thick like the Galactic thick
disk and flared.
Therefore, the LMC is a disk galaxy
that has a halo, but not a bulge,
and in this regard it is like M33 
(van den Bergh 1991).

It is not known
how stellar halos form, or how they relate to the other
spheroidal components of galaxies
(e.g., Wyse \& Gilmore 1988).
Stellar halos are minor constituents of disk galaxies, and
thus they can be difficult to detect.
The proximity of the LMC 
therefore affords a unique opportunity to
study an extragalactic stellar halo.  
How does the LMC's stellar halo compare to the 
Milky Way's stellar halo?
This question illustrates the point that
the LMC is a benchmark for understanding galaxy formation and structure,
and hence also why measuring
the basic parameters of the LMC's stellar halo is important.
An accurate model of the LMC's
stellar halo also provides a basis for interpreting microlensing results.
In fact
the stellar halo is a new population unknown 
in detail to any prior
discussion of LMC microlensing, although
its possible existence has been considered
(e.g., Alcock et al.~2000).   

Measurements of the surface density of RR Lyrae stars in the LMC
have a venerable history of being used to study
the LMC's stellar halo
(Kinman et al.~1991).
RR Lyrae stars are also used to map the structure of
the Galactic stellar halo
(e.g., Wetterer \& McGraw 1996), and thus their distribution 
in the LMC can be compared directly to these Galactic data.
The starting point for the present work is Fig.~12
of Alcock et al.~(2000b) where the 
radial profile of RR Lyrae star counts in the LMC
based on data
from 16 MACHO survey fields was presented.
There are several reasons why that analysis 
should be revisited.
First, the primary conclusion of Alcock et al.~(2000b) regarding
the star-count data was the good fit of an exponential disk model;
halo models were not thoroughly compared.
In addition, new and more accurate catalogs of RR Lyrae stars
in 30 MACHO survey fields have since been compiled (inclusive of
the 16 noted above), and artificial
star tests have now also yielded statistical completeness corrections.
Last, a new catalog of
RR Lyrae stars from the OGLE~II survey
is available
(Soszy\'{n}ski et al.~2003), and these data can 
be analyzed together with the MACHO~data.
The results of this
analysis are used
to estimate the total mass and mass-to-light ratio of the LMC,
and to make predictions about LMC microlensing.

\section{ Properties of the LMC Spheroid }

One way to describe the LMC stellar halo
is with King's (1962) model for an 
isothermal system of particles modified by an 
external tidal field.
The approximate formula
for fits to surface brightness data is:
\begin{equation}
 S(R) = S_{0} \left[
 \left(\ 1 + \left(R/a\right)^2\ \right)^{-1/2} \  - \  
 \left(\ 1 + \left(R_t/a\right)^2\ \right)^{-1/2} \    \right]^2
\end{equation}
where $R$ is the projected radius,
and the parameters are the central normalization ($S_{0}$),
the core radius ($a$), and the tidal radius ($R_t$).
For $R_t >> a$, Eqn.~(1) simplifies to
the ``Hubble-Reynolds'' profile.
A good approximation for the density profile of a King model
is the ``modified Hubble'' function:
\begin{equation}
\rho(r) = \rho_{0}
\left[ \ 1 + \left(r/a\right)^2 \ \right]^{-3/2}
\end{equation} 
where 
$r$ is the spherical radial coordinate.
The projection of the
modified Hubble function is exactly the Hubble-Reynolds profile,
and thus $S_{0} =  2 \rho_{0} a$ for $R_t >> a$.
The central density ($\rho_0$) in mass units
is related to the velocity dispersion ($\sigma$) by
$\rho_0 = 9 \sigma^2 / 4 \pi G a^2$ (Rood et al.~1972).

The catalog of RR Lyrae stars used by Alcock et al.~(2000b)
is known to have serious systematic errors
in the number of stars in some fields (e.g., Alcock et al.~2003).
The most accurate catalog of RR Lyrae stars
in the MACHO database has not yet been published, but was
provided by K.~Cook (private communication) 
for the present investigation.
Cook's summary\footnote{Data available by request to the author.}
refers only to stars
that pulsate in the fundamental mode (type RR0 or RRab).  
These are the easiest to detect due to their large pulsation
amplitudes, and they are traditionally used to study the
stellar halos. 
The completeness of the MACHO database is now also better understood
because of artificial star tests that were part of
the microlensing detection efficiency calculation
(Alcock et al.~2001).  
Unpublished plots of the fraction of artificial stars recovered 
in the MACHO database as a function
of input magnitude were constructed by T.~Vandehei 
(private communication) for 6 
example ``chunks'' of MACHO image data\footnote{In fact 
the results of these tests are not reflected in the
luminosity functions published by Alcock et al.~(2001),
but the results of that work are not affected by the omission.}.
The completeness fraction of interest has been read off of 
Vandehei's diagram at $V$ = 19.4 mag, which represents a
typical LMC RR Lyrae star.  
As expected,
the fraction of recovered $V$=19.4 stars ($f$)
is correlated with the density of detected objects 
in each example sub-field (Alcock et al.~2001).
The result of a linear fit is
$f$ = 1.37$\cdot$N/N$^\prime$
= $1.0 - 0.00073 \cdot O$, where $O$
is detected objects per arcmin$^2$
(Alcock et al.~2001, see their Table 2),
N is the number of RR0 stars, and N$^\prime$ is the estimated total number
of all types of RR Lyrae stars.
This approximate correction 
has an uncertainty of 7\% based on the standard deviation
of the fit to the artificial data.

A new catalog of
LMC RR Lyrae stars 
has also recently been published by the OGLE~II survey
(Soszy\'{n}ski et al.~2003).
This catalog includes RR Lyrae stars 
in all pulsation modes, not just the RR0 stars.
Soszy\'{n}ski et al.~(2003) suggest that the completeness 
of their catalog is 95\%,
but this is not consistent with the calibration 
obtained below as a free parameter in the profile fit.
In order to incorporate these data,
the type RR0 stars from OGLE~II fields 3--9 are 
binned into 55 equal-area (96.8 arcmin$^2$) regions that blanket the
central bar.

I adopt a ratio of 1.37 for the total number 
of RR Lyrae stars to the number of RR0 stars
in stellar halos from
Kinman et al.~(1991).  This factor is included in the
completeness correction formula above.
Note, however, that the Soszy\'{n}ski et al.~(2003) catalog 
implies a ratio of 1.30 for the LMC.
The estimate by Kinman et al.~(1991) is preferred because it
is based on Galactic data,
which may be more accurate.
In particular, the lowest-amplitude 
RR Lyrae stars are mostly not of type RR0,
and thus it is possible that relatively fewer of these
have been detected in the LMC,
which would tend to lower the LMC ratio.
In any case, the main conclusions of this work are
not significantly affected by this assumption.

The King-model fits are performed on the
MACHO and OGLE~II surface brightness data 
($V_0$ mag/arcsec$^2$)
presented in Fig.~1.
It is assumed that every RR Lyrae star represents
a luminosity of 6730 $L_{V,\odot}$ in halo stars
(Kinman et al.~1991).
I adopt a standard distance modulus of 18.5 mag (50.1 kpc),
and a tidal radius of
$R_t$ = 15 kpc 
(van der Marel et al.~2002). 
The adopted value of $R_t$
does not significantly affect the derived parameters
(see \S2.1).
In all there are 85 nearly independent measurements and
5 parameters to fit.
The free parameters are the center of the halo,
$\alpha$ and $\delta$ (J2000), $a$ and $S_0$
from Eqn.~(1), and the fractional incompleteness 
of the OGLE~II data.  Inspection of Fig.~1 shows that
the majority of MACHO and OGLE~II data points are on
different lines-of-sight,
and are thus independent.  The analysis
is simplified by assuming that all of the points are
independent.
An additional 7\% error associated with the completeness corrections
is added in quadrature with the $\sqrt N$ errors,
and these are the error bars shown in the main panel of Fig.~1.
The best fit has 
$\chi^2$/dof = 1.19~(dof=79) and parameters:
$\alpha = 5^{\rm h} \ 21.1 \pm 0.8\,^{\rm m}$,
$\delta = -69^{\rm \circ} \ 45 \pm 6\arcmin$,
$a = 1.42 \pm 0.12$ kpc, and
$S_0 = 23.13 \pm 0.07$ mag/arcsec$^2$.
This model is the solid line.
The $\chi^2$/dof obtained
lends strong support to the accuracy
of the derived and adopted parameters.
Note that
the definition $S_0$ in Eqn.~1 is not the same as
the maximum central surface brightness of the halo,
which reaches only 23.35 mag/arcsec$^2$.
The ``incompleteness parameter'' used
in the fit is a way to scale the
OGLE~II data to best match the 
profile of the completeness-corrected MACHO data,
and on this basis the OGLE~II data
employed are only about 80\% complete.
Finally, inspection of a map of the fit residuals (not shown)
reveals no indication of ellipticity.
A main result of this calculation 
is the estimate of the core radius.

The center of the distribution of 
RR Lyrae stars agrees well
with the infrared and optical centers of the LMC bar:
$\alpha$ = $5^{\rm h} \ 25.0 \pm 0.1\,^{\rm m}$,
$\delta$ = $-69^{\rm \circ} \ 47 \pm 1\arcmin$ 
(van der Marel 2001), and
$\alpha$ = $5^{\rm h} \ 23.6\,^{\rm m}$, 
$\delta$ = $-69^{\rm \circ} \ 44\arcmin$ 
(de~Vaucouleurs \& Freeman 1973), respectively.
For comparison, Soszy\'{n}ski et al.~(2003) report
$\alpha$~=~$5^{\rm h} \ 22.9\,^{\rm m}$,
$\delta$ = $-69^{\rm \circ} \ 39\arcmin$
for the center of the RR Lyrae distribution
in the OGLE~II catalog.  
Overall, these estimates suggest 
that the LMC center is known to an accuracy of
about 10$\arcmin$.
Note that the RR Lyrae center is 
grossly inconsistent the kinematic center
of the \ion{H}{1} gas,
but it is consistent
with the kinematic center of the carbon stars
(van der Marel et al.~2002).

\subsection{Tidal Radius}

Before microlensing surveys existed,
surface density measurements
of LMC RR Lyrae stars were made 
by blinking photographic plate images.
These data, although obtained by primitive methods,
provide a direct
test of the accuracy of the halo model derived above.
In addition, the fields surveyed with photographic plates
are located farther from the center of the LMC than
the now available microlensing survey data,
and thus they should be useful 
to constrain the tidal radius,
and to test if the halo is spherical at these radii.
Of course the situation might not be so simple.
For example, the possible existence
of extra-tidal stars near the LMC's periphery
could complicate the interpretation of the tidal radius.

The photographic RR0 star-count data
summarized by Kinman et al.~(1991)
are reproduced
in Table~1, which lists the field name (associated
with a globular cluster in each field), the radial distance 
from LMC center, the area surveyed, and the
observed number of RR0 stars.
The numbers of RR0 stars 
expected according to the spherical halo model 
are listed for easy comparison.
These data are also
compared 
in the inset panel of Fig.~1.
Significant discrepancies are obvious.  The worst case
is the NGC\,1783 field where the observed number of RR0 stars
is probably only about 50\% complete.
However, the agreement 
is fair in the two outermost fields (Reticulum \& NGC\,1841).
Inspection of the images of these
fields underlines the fact that they are very empty, and thus
confusion is not an issue,
and that the RR0 stars are detected with reasonably high signal-to-noise.
Therefore, the photographic data should be interpreted with caution,
but perhaps not dismissed entirely.
Finally, however, an attempt to fit Eqn.~(1) to these photographic data
yields no better constraint on the tidal
radius than the current best kinematic estimate 
of $15 \pm 4.5$ kpc
(van der Marel et al.~2002).

\section{Implications for the Dark Halo}

The new model of the LMC stellar halo yields
information about the total mass of the LMC complementary to what
is already known from the rotation curve.  
The most accurate estimate of the maximum circular velocity 
of the LMC disk
is based on carbon stars (van der Marel et al.~2002): 
$V_C = 65 \pm 16$ km s$^{-1}$ at the last measured
radius of 8.9 kpc.
Extrapolating out to the tidal radius 
implies a total mass of $1.47 \pm 0.85 \times10^{10} M_{\odot}$.
The estimated total $V$-band luminosity of the LMC
is $3.0 \times 10^9 L_V$ in Solar units, and hence
the global mass-to-light ratio is $M/L_V = 4.9 \pm 2.8$
(van der Marel et al.~2002; see also Alves \& Nelson 2000).

If the stars in the core 
of the LMC's stellar halo have a Maxwellian
velocity distribution, then the
expected velocity dispersion 
is $\sigma = V_C/\sqrt{2} = 46 \pm 11$ km~s$^{-1}$.
This is in good agreement with 
$\sigma = 53~\pm~10$ km~s$^{-1}$ measured
by Minniti et al.~(2003).
Adopting the measured estimates of $\sigma$ and $a$,
King's core-fitting method yields 
a central mass density 
$\rho_0 = 2.3 \pm 1.0 \times 10^8 \, M_{\odot}$ kpc$^{-3}$.
The assumptions of King's method are that
mass follows light, 
that the orbits of the RR Lyrae stars are isotropic, and 
that the LMC is in equilibrium.
The central surface brightness of the LMC 
is 20.56 mag/arcsec$^2$ (Bothun \& Thompson 1988),
which therefore implies $M/L_V = 5.3 \pm 2.1$
for $S_{0} =  2 \rho_{0} a$.
For the total $V$-band luminosity given above,
the predicted total mass 
of the LMC is $1.6 \pm 0.6 \times10^{10} M_{\odot}$.
The error bars on the estimated
mass-to-light ratio and total mass are dominated 
by the uncertainty of $\sigma$.

\section{The Stellar Halo Resembles the Disk}

Instead of a spherical halo model, 
an exponential disk model 
fit to the data in \S2 
yields a radial scale length of
$\lambda = 1.47 \pm 0.08$ kpc, a maximum central
surface brightness of $23.15 \pm 0.07$~mag/arcsec$^2$,
and $\chi^2$/dof = 1.19~(dof=82).
This model is the dotted line in Fig.~1, and it is
statistically indistinguishable from the best-fit halo model.
(The center and incompleteness of the OGLE~II data are
fixed to the values found above, but the results are the
same if this assumption is relaxed.)
An exponential is known to usually provide a good fit to 
the core of an isothermal, as found here.

Although the LMC's halo and disk are kinematically different,
the shape of the surface density profile of the RR Lyrae stars
is remarkably similar to that of the young disk.
For example, the best-fit exponential
scale length for the RR Lyrae stars is very close to
the scale length of the LMC's blue light: 1.46~kpc 
(Bothun \& Thompson 1988).  For comparison,
the Milky Way's stellar halo as traced by RR Lyrae stars
does not look like an exponential disk; it has a power-law
density profile $\rho(r) \propto r^{-3.024}$ (Wetterer \& McGraw 1996)
from 0.6 to 80 kpc, which projects to a surface density
profile that is obviously different from exponential models
that best represent the Milky Way's disk.

\section{Implications for Microlensing }

The basic parameters of the
LMC's stellar halo were unknown to any prior
discussion of LMC microlensing 
(e.g., Salati et al.~1999; Gyuk et al.~2000).
The following approximate calculation 
is the first to use a modified Hubble density model.
The optical depth
for a thin inclined disk embedded in a
distribution of lenses is given by 
(Gould 1993;  Guidice, Mollerach \& Roulet 1994):
\begin{equation}
\tau = \int\limits^{D_{OS}}\limits_0  \frac{4\pi G\rho (r)}{c^2}
\ell \left(1 - \frac{\ell }{D_{OS}}\right) d\ell
\ \approx \
\frac{4\pi G\rho_{\circ} a^3}{c^2} 
\int\limits^{\ell_{max}}\limits_0 \frac{ \ell ~d\ell }{
\left[\ell ^2 + k_1\ell + k_2 \right]^{3/2} }
\end{equation}
where $r$ is the spherical radial coordinate,
$\rho (r)$ is the density of lenses,
$D_{OS}$ is the distance between the observer and source stars,
$\ell$ is the line of sight, and
the density $\rho(r)$ is from Eqn.~(2).
The approximation is to drop the term $\ell/D_{OS}$
because the LMC's tidal radius
is a factor of a few smaller than the LMC's distance.
The constants in the denominator are:
$k_1 = -2b\cos\,\phi\tan\,i$ and
$k_2 = b^2\cos^2\phi\tan^2i + a^2 + b^2$,
where $\phi$ is position angle relative to the 
far minor axis of the disk,
the impact parameter is $b$,
and the disk has an inclination angle $i$.
The integration limit is
$\ell_{max} = b\cos\,\phi\tan\,i + \left(R^2_t - b^2\right)^{1/2}$.
Dwight (1961) solves the integral (see Eqn.~380.013), and thus:
\begin{equation}
\tau \approx \frac{4\pi G\rho_{\circ} a^3}{c^2}\left[
\frac{\left(R^2_t - b^2\right)^{1/2}}{\left(R^2_t + a^2\right)^{1/2}}
\frac{\left(b~\cos~\phi ~\tan~i\right)}{\left(a^2 + b^2\right)} \\
\ + \ 
\frac{\left(\ a^2 + b^2 + b^2 \cos^2\phi ~\tan^2i \ \right)^{1/2}}{
\left(a^2 + b^2\right)}
\ - \  \frac{1}{\left(R^2_t + a^2\right)^{1/2}}
\right]
\end{equation}
The first term in Eqn.~(4) is an odd power of $\cos \phi$, 
which yields the highest
values on the far
side of the inclined disk.

The LMC disk is generally known to contain 
interstellar dust, 
and this opacity will tend
to suppress the rate of disk-on-halo lensing relative to
halo-on-disk lensing.  Therefore, the
existence of the LMC stellar halo implies a new local maximum
in the LMC optical depth map whose location depends in part
on the parameters of the halo as given by Eqn.~(4).
A ratio of $M/L_V$ = 1.6 is typical of
stellar halos (Kinman et al.~1991), and this implies
$\rho_0 =  6.3 \times 10^6 \, M_{\odot}$ kpc$^{-3}$ 
for the LMC stellar halo following \S2.
Adopting a disk inclination of 34.7$^{\circ}$
and a line-of-nodes at
position angle $\sim$130$^{\circ}$ measured East of North
(van der Marel et al.~2002),
Eqn.~(4) predicts a
maximum of $\tau = 9 \times 10^{-9}$ at a distance of
0.95 degrees from center at position angle $\sim$220$^{\circ}$.
This is nearest to microlensing events 4, 5, 22, and 23 from 
Alcock et al.~(2000; their Table~8).
Excluding event 5, which is caused by
a lens in the Galaxy, and event 22, which is 
a peculiar supernova, events 4 and 23 account for
an optical depth of $1.2\times10^{-8}$.
Therefore, based on location and approximate contribution
to the observed optical depth, it is possible
that MACHO events LMC-4 and LMC-23 are
halo-on-disk stellar self-lensing.
The remaining events in the MACHO sample
(excluding 5, 22, and the 
self-lensing event LMC-14; Alcock et al.~2001b)
account for an optical depth of $8.2\times10^{-8}$,
which is not explained by stellar lenses in the LMC or Galaxy.

\section{Conclusion}

A King model fit to RR Lyrae star-count data in the LMC
yields the center of their distribution at
$\alpha$ = $5^{\rm h} \ 21.1 \pm 0.8\,^{\rm m}$,
$\delta$ = $-69^{\rm \circ} \ 45 \pm 6\arcmin$ (J2000)
and a core radius of $1.42 \pm 0.12$ kpc.
As a check, 
the predicted numbers of halo RR Lyrae stars near the LMC periphery
are compared with counts based on photographic plates.
However, these data require a cautious interpretation.
King's core-fitting method implies
a mass-to-light ratio $M/L_V = 5.3 \pm 2.1$
and total LMC mass $1.6 \pm 0.6 \times 10^{10} M_{\odot}$.
It is noted that the shape of 
the surface density profile of the LMC's halo
is very similar to that of its young disk, which is not the
case for the Galactic disk and halo system.
Finally, the stellar halo is predicted to cause a local maximum
of microlensing optical depth large enough
to possibly
explain MACHO events LMC-4 and LMC-23,
but not all of the observed events.
Further observations of RR Lyrae stars that might
test the spherical LMC halo model include more 
surface-density data near the tidal 
radius, more kinematic data, and accurate distances on 
multiple lines of sight.

\acknowledgments

I thank Ted Baltz, Sergei Nikolaev, Kem Cook, and Dante Minniti
for their many helpful discussions about
early drafts of this paper, 
and Scott Gaudi
for his thoughtful referee reports.
This paper utilizes data obtained by the MACHO Project, 
jointly funded by the US Department of Energy through the 
University of California (UC), Lawrence Livermore National Laboratory 
under contract No. W-7405-Eng-48, 
by the National Science Foundation through the UC Center for Particle 
Astrophysics under agreement 
AST-8809616, and by the Mount Stromlo and Siding Spring Observatory, 
part of the Australian National University.


\clearpage

\begin{deluxetable}{lllll}
\tablecaption{RR Lyraes from Photographic Surveys}
\tablewidth{0pt}
\tablehead{
\colhead{Field} &
\colhead{$\log$ R} &
\colhead{A} &
\colhead{$N_O$ } &
\colhead{$N_P$ } 
}
\startdata
NGC 1783  & 0.63  & 1.3 & 61  & 144   \\
NGC 2210  & 0.65  & 0.64 & 40 & 64  \\
NGC 1466  & 0.91  & 0.6 & 3   & 10  \\
NGC 2257  & 0.93  & 0.6 & 15  & 8  \\
Reticulum & 1.08  & 0.6 & 1   & 1.5 \\
NGC 1841  & 1.16  & 0.6 & 2   & 0.2  \\
\enddata
\tablecomments{This table refers only to type RR0 (=RRab) stars. 
$N_O$ is observed number from Kinman et al.~(1991); 
A is surveyed area in deg.$^2$; R is radius from center in deg.;
$N_P$ is the number predicted by the halo model (see text).
}
\end{deluxetable}

\clearpage

\begin{figure}
\plotone{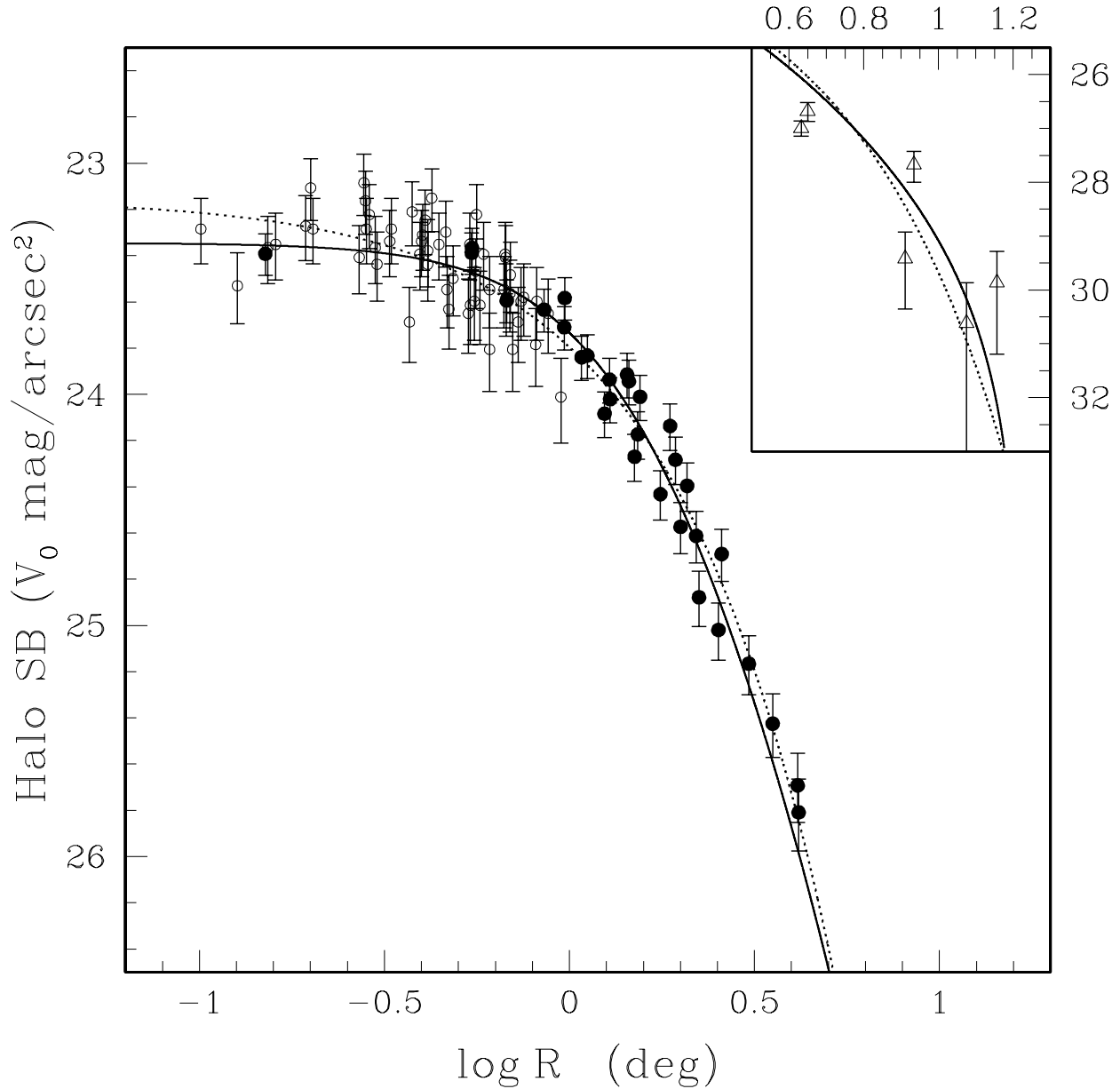}
\caption{Surface brightness of the LMC stellar halo calculated
from RR Lyrae star counts.  In the main panel,
the error bars are $\sqrt N$ plus a 7\% contribution
associated with the statistical completeness corrections.
The solid line is the best-fit halo model, and the dashed line is
the best-fit exponential disk model. The inset panel shows
photographic data (Table~1) for fields at large
radial distance; these error bars are $\sqrt N$. }
\end{figure}

\end{document}